\documentclass[12pt, letterpaper]{article}
 \pdfoutput=1
\usepackage{amssymb}
\usepackage{amsmath}
\usepackage{amstext}
\usepackage{graphicx,epsfig}
\usepackage{epsfig}
\usepackage{verbatim} 
\usepackage{fancybox}

\usepackage[body={17.5cm, 22cm},right=2.2cm]{geometry}

\newcommand{\be}{\begin{equation}}
\newcommand{\ee}{\end{equation}}
\newcommand{\bea}{\begin{eqnarray}}
\newcommand{\eea}{\end{eqnarray}}
\newcommand{\beas}{\begin{eqnarray*}}
\newcommand{\eeas}{\end{eqnarray*}}

\newlength{\fdagwidth}
\newlength{\diagupwidth}
\newlength{\stepback}

\newcommand{\sfrac}[2]{{\textstyle\frac{#1}{#2}}}
\newcommand\di{\partial}

\begin{document}

\begin{center}

~

\vspace{1.cm}

{\LARGE \bf{
Superluminality in DGP
}}\\[1.5cm]
{\large Kurt Hinterbichler$^{\rm a}$, Alberto Nicolis$^{\rm a}$, Massimo Porrati$^{\rm b}$}
\\[.8cm]

\vspace{.2cm}
{\small \textit{$^{\rm a}$ 
Department of Physics and ISCAP\\
 Columbia University, 
New York, NY 10027, USA }}

\vspace{.2cm}
{\small \textit{$^{\rm b}$ 
Center for Cosmology and Particle Physics\\
Department of Physics, 
New York University\\
4 Washington Place, New York, NY 10003, USA}}

\end{center}

\vspace{.8cm}
\begin{abstract}

We reconsider the issue of superluminal propagation in the DGP model of 
infrared 
modified gravity. Superluminality was argued to exist in certain otherwise 
physical backgrounds by using a particular, physically relevant scaling 
limit of the theory.

In this paper, we exhibit explicit five-dimensional solutions of the full 
theory that are stable against small fluctuations and that indeed support 
superluminal excitations. The scaling limit is neither needed nor invoked 
in deriving the solutions or in the analysis of its small fluctuations. 
To be certain that the superluminality found here is physical, we analyze the 
retarded Green's function of the scalar excitations, finding that it is 
causal and stable, but has support on a widened light-cone.
We propose to use absence of superluminal propagation as a method to constrain
the parameters of the DGP model. As a first application of the method, we find 
that whenever the 4D energy density is a pure cosmological constant and
a hierarchy of scales exists between the 4D and 5D Planck masses, 
superluminal propagation unavoidably occurs.

\end{abstract}


\section*{Introduction}
\ \ \ \ \
The possibility of modifying general relativity at large distances is both
theoretically fascinating and of potentially great import for cosmology. 
The theoretical challenge stems from decades of negative results about the
very possibility of changing the long distance behavior of gravity. Various
roadblocks stand in the way of such a modification. The first is the 
celebrated van Dam-Veltman-Zakharov discontinuity~\cite{vdvz}, which is due to
the fact that in flat space a massive spin-2 field propagates a zero helicity
state. In the massless limit, that helicity becomes a scalar that couples 
with gravitational
strength to the trace of the stress-energy tensor of matter. Such coupling 
changes the GR prediction of, say, the bending of light by massive objects by
a finite amount incompatible with experiment. This is to say that the massless
limit of a theory of massive gravity is not GR but rather
a Brans-Dicke theory with Brans-Dicke
parameter $\omega=O(1)$. Current solar-system observations imply 
$\omega> 40,000$. More theoretical but potentially 
worse problems 
are the existence of ghosts around Schwarzschild-like solutions~\cite{dr}, the
absence of a lower bound on the energy of small fluctuations around generic
backgrounds~\cite{bd}, and the existence of an extra propagating degree 
of freedom, beyond linear order, in generic nontrivial backgrounds 
(also in~\cite{bd}).
All these problems are related, and call for a more sophisticated way of 
changing gravity at large distances. The DGP model~\cite{dgp} provides at least
in part a cure for some of the problems just mentioned. In DGP, gravity is 
four dimensional below a certain distance $L$, and five dimensional above $L$. This
distance is given by the ratio of the 5D Newton's constant 
over the 4D Newton's constant,  $L=G_5/G_4$.\footnote{Equivalently 
$L=M_4^2/M_5^3$, where $M_4$ is the 4D Planck mass while $M_5$ is the
Planck mass in 5D.}
Thus, the familiar $r^{-2}$ behavior of Newton's law becomes a weaker 
$r^{-3}$ at large 
distances; empirically, $L={\cal O}(R_{\rm Hubble})$.
While still suffering~\cite{LPR}
from the severe strong-coupling problem which also plagues massive 
gravity~\cite{ahgs} at the quantum level, the DGP model 
does not propagate ghosts on 
asymptotically flat backgrounds~\cite{nr}. 
Ref.~\cite{nr} also 
shows that the crucial aspect of the DGP dynamics, namely the
decoupling of the scalar degree of freedom around a massive source, holds in a 
decoupling limit in which the 4D Planck mass $M_4\rightarrow \infty$, 
while the quantum DGP scale $\Lambda_3= ({M_4/L^2})^{1/3}$ is kept constant.
In this limit, gravity around a source of (fixed) Schwarzschild radius $R_S$ is
in the linear regime whenever $r\gg R_S$, with the exception of a self-interacting scalar
mode, $\pi$, that couples with gravitational strength to the trace of the stress-energy
tensor. The strong self-coupling of the scalar causes it to decouple at distances
$r \ll (R_S L^2)^{1/3}$~\cite{v,dvali,P,nr}, and to propagate superluminally in the range
$ R_S \ll r \ll (R_S L^2)^{1/3}$~\cite{nima}. A recent work~\cite{nrt} shows that
this pathology of superluminality is generic in the decoupling limit.

A subtle aspect of the decoupling limit is that it washes away the non-local
part of the scalar propagator. In DGP, the degree of freedom that becomes the
scalar of the decoupling limit is one of the helicities of the graviton, whose
propagator is non-local. The DGP graviton is indeed a resonance, not a stable
particle (it can decay from the 4D brane into the 5D bulk). Equivalently, its
propagator (better, its retarded Green's function) does not have poles on the
physical sheet of the complex frequency plane, but only cuts along the 
real axis. Moreover, the kinematic region of interest for studying locality 
is precisely that where the non-local part of the propagator cannot be 
neglected. Another possible complaint about the conclusions reached 
in~\cite{nima, nrt} is that the background on which the scalar mode propagates is
only a solution of the decoupled equations and not of the full DGP model.
Finally, in a model such as DGP in which 4D non-localities stemming from the
integration of the 5D bulk physics exist, one must exercise care in distinguishing
true superluminal propagation from non-localities introduced by an incorrect 
definition of the sources~\cite{dntv}. 

To respond to all these doubts, after a lightning review of the DGP equations in section 2, we will review some of its exact solutions in section 3; the time independent ones of~\cite{LPR} and the time-dependent ones of~\cite{gkmp}.  In particular, we compute the kinetic term of the scalar DGP mode, $\pi$, in the backgrounds found earlier.
From that computation we find a speed of propagation of signals $c_\pi$ 
greater than $c$,
thereby confirming the results of~\cite{nr}. Section 4 is devoted to
studying the issue of non-locality of $\pi$ 
and of a proper definition of its sources. To be
sure that $\pi$ indeed propagates superluminally, we study its retarded
Green's function {\em without} resorting to the local approximation. Our 
analysis confirms that the Green's function is indeed nonzero inside an 
enlarged light cone, determined by $c_\pi$. Equally importantly, we find
that the Green's function does not have instabilities growing exponentially in
time, in accordance with the analysis done in the decoupling (or local) limit.

In section 3 we tackle a di?erent problem.  We 
examine whether a range of parameters exists, for
which the DGP model admits only solutions without superluminal propagation
of signals. In other words, 
we try to use special-relativity causality to 
constrain the parameter space of DGP. We find that, at least in the case that
the 4D brane energy is a pure cosmological constant, no range of parameter is
safe. This may point to an upper bound on $L$ (the DGP length scale) or on
the horizon radius in any relativistically invariant, causal completion of the model, such as those
provided by D-brane or other constructions in string theory.

\section{The model, and our goal}
\ \ \ \ \
DGP is a 5D model where gravity propagates throughout an infinite 5D bulk, and  matter fields are confined to a 4D boundary. The action for gravity at lowest order in the derivative expansion is a bulk Einstein-Hilbert term and a boundary one, generically with two different Planck masses $M_5$, $M_4$ (plus a suitable Gibbons-Hawking term). Even if we set the gravitational boundary action to zero at tree-level, matter loops will generically generate a localized 4D Einstein-Hilbert term. The phenomenologically interesting regime is $M_5 \ll M_4$.

The DGP equations are the Einstein equations in the bulk,
\footnote{We are using the mostly plus signature for the metric. Also we define the $D$-dimensional Planck mass as $M_D^{D-2} = 1/8\pi G_D $. }
\be \label{5deinstein} 
G^{(5)}_{MN}=0 \; ,
\ee
along with the Israel junction conditions on the brane (with $\mathbb{Z}_2$ symmetry across the brane),
\be \label{junctioncondition} 
M_4^2 G_{\mu\nu}-2 M_5^3\left(K_{\mu\nu}-g_{\mu\nu}K\right)=T_{\mu\nu}
\ee 
(a boundary is equivalent to a brane with $\mathbb{Z}_2$ symmetry).
$K_{\mu\nu}$ is computed with the normal vector pointing into the bulk.  Note that we still have $\nabla_\mu T^{\mu\nu}=0$, by virtue of $\nabla^\mu\left(K_{\mu\nu}-g_{\mu\nu}K\right)=0$, coming from the momentum constraint implicit in (\ref{5deinstein}).

Ref.~\cite{LPR} studied small fluctuations about a generic curved solution $\bar g_{MN}$. Since the stress-energy tensor is localized on the boundary, it is interesting to `integrate out' the bulk modes and rewrite the theory as a 4D theory. Namely, one can solve the linearized 5D Einstein equations for given sources and given 4D boundary conditions, plug the solution back into the action, and thus get an effective four dimensional action that only involves  the sources and the 4D field configuration on the boundary. The downside of doing so is that such an effective action is non-local, because massless degrees of freedom have been integrated out. Remarkably, at short-distances the effective action is approximately local---hence the statement that in the model gravity looks four-dimensional at short distances.
Indeed the resulting quadratic action for small fluctuations at distances much shorter than the solution's curvature radius and than the critical DGP length-scale is \cite{LPR}
\bea \label{effectiveS}
S^{\rm 4D}_{\rm eff} & \simeq & \int \! d^4 x \sqrt{- \bar g} \Big[ M_4^2 \big( \sfrac12  h^{\mu\nu} \, \bar \Box h_{\mu\nu} - \sfrac14 h \, \bar \Box h \big) + \sfrac12 h_{\mu\nu} \, \delta T^{\mu\nu} \nonumber  \\
&& + \;  {3} m^2 M_4^2 \,  \pi \,  \bar \Box \pi  +   2m M_4^2 \, \pi \big( \bar K_{\mu\nu} - \bar g_{\mu\nu} \bar K \big) \bar \nabla^\mu \bar \nabla^\nu \pi 
+ \sfrac12 m \, \pi \, \delta T^\mu {}_\mu 
\Big] \; ,
\eea
where $m \equiv 1/L_{\rm DGP} = 2M_5^3/M_4^2$ is the DGP mass scale, all `barred' quantities are computed on the solution $\bar g_{\mu\nu}$, and we are implicitly assuming de Donder gauge for $h_{\mu\nu}$: $\bar \nabla^\mu (h_{\mu\nu} - \sfrac12 \bar g_{\mu\nu} \, h) = 0$. $\pi$ is a scalar degree of freedom, which can be thought of as a `brane bending' mode. Indeed in one gauge it measures fluctuations in the position of the brane along the fifth dimension. In another gauge it is a component of the five-dimensional metric. The gauge-invariant, physical statement is that it couples directly to $T_{\mu\nu}$, and thus mediates a physically measurable force.\footnote{To get to (\ref{effectiveS}), $\pi$ has been demixed from the metric perturbation $h_{\mu\nu}$ \cite{LPR}. This is the origin of the coupling $\pi T^\mu {}_\mu$. Equivalently, matter couples universally to the `Jordan-frame' metric perturbation $\hat h_{\mu\nu} = h_{\mu\nu} + m \pi\,  \bar g_{\mu\nu}$.
}
The non-local contributions alluded to above, are corrections to the above action schematically of the form
\be
M_4^2 m \, h \, \sqrt{- \bar \Box} \, h \; , \qquad M_4^2 m^3 \, \pi \, \sqrt{- \bar \Box} \, \pi \; .
\ee
At high momenta, $\di \gg m$, their effect is negligible with respect to the local kinetic terms we explicitly kept. For the moment, we will concentrate on the local part of the quadratic action, and come back to the interpretation (and a more careful definition) of these non-localities in Sect.~\ref{green}.

We see that the $\pi$ kinetic action---thus its propagator---depends on the extrinsic curvature of the solution.
In the following, we will look for 5D background solutions where this extrinsic curvature correction makes $\pi$ excitations superluminal while keeping their energy positive.  That is, we will look for stable solutions with superluminal excitations.

An important remark is in order. One might think that, even if we find superluminal excitations of curved solutions, these excitations will simply correspond to signals connecting two points $A$ and $B$ on the curved boundary by taking a straight shortcut through the bulk. In this case our 4D superluminality would be fake---no signal would exit the true 5D light-cone. This however cannot be the case, for two reasons. First, the extrinsic curvature correction to $\pi$'s kinetic action above is {\em local}: it can give a finite shift in the local propagation velocity of $\pi$, for arbitrarily close $A$ and $B$. The shortcut effect, instead, is only efficient for $A$ and $B$ far from each other, so that the curvature of the boundary becomes visible. The shortcut fractional correction to the average velocity goes to zero if we bring $A$ and $B$ closer and closer. In other words, a shortcut effect would correspond to a non-local term in the effective 4D Lagrangian.
Second, linear $\pi$ excitations correspond to  pure gauge deformations of the bulk geometry \cite{LPR}. Therefore, when we excite $\pi$ no physical information is traveling through the bulk. We can choose a gauge such that the bulk is unperturbed, and the geometric excitation is manifestly localized on the boundary.


\section{Stable solutions with superluminal $\pi$}
\ \ \ \ \
In order to {\em (a)} find bulk solutions, and {\em (b)} study the kinetic action for $\pi$ about them, it is convenient to focus on configurations with a high degree of symmetry. One possibility is to consider spatially flat cosmological solutions, for which the bulk geometry is exactly flat.
There, however, one can show that imposing the null energy condition (NEC) for $T_{\mu\nu}$ and stability under $\pi$ fluctuations, automatically implies {\em sub}-luminality of $\pi$  \cite{HN}.
The next-to-simplest possibility is solutions that are spherically symmetric (in five dimensions), with the brane curled up as a 3-sphere. In such a case, Birkhoff's theorem forces the bulk metric to be static, and of the 5D Schwarzschild form:
\be 
ds_{(5)}^2=-f^2(r)dt^2+{1\over f^2(r)}dr^2+r^2d\Omega_{3}^2 \; , \qquad f(r)=\sqrt{1-{\mu\over r^2}} 
\ee
(here and in the following we use $f^2$ to denote what is usually called $f$).
The parameter $\mu$ is related to the solution's 5D ADM mass by
\be 
M_{\rm ADM} = 3\pi^2M_5^3\mu \; .
\ee

\subsection{Time independent spherical solutions}\label{static}
\ \ \ \ \
We first look for solutions with a static spherical brane located at some fixed value of $r$. This case has been considered in \cite{LPR}. From the 4D viewpoint, the brane is a static Einstein Universe. The 4D induced metric is
\be
g_{\mu\nu} dx^\mu dx^\nu = - f^2(r) dt^2 + r^2 d\Omega_3 ^2 \; ,
\ee
and the extrinsic curvature is
\be
K_{\mu\nu} =\frac12 f \, \partial_r g_{\mu\nu} \; .
\ee
We have chosen the bulk to lie outside the brane, so that it is asymptotically flat and free of horizons and singularities.  Then the DGP equations (\ref{junctioncondition}) read \cite{LPR}
\bea
\frac{1}{(mr)^2} -\frac{f}{mr} & = & \frac{1}{3M_4^2m^2} \rho, \label{E00} \\
\frac{1}{(mr)^2} -\frac{1}{mr} \left( f+\frac{1}{f} \right) & = & - \frac{1}{M_4^2m^2} p, \label{Eii}
\eea 
where $m \equiv 2M_5^3/M_4^2 $ is the DGP mass parameter, $r$ is the brane position, $f$ is computed at $r$, and $\rho$ and $p$ are the 4D energy density and pressure.  Note that the energy-momentum conservation equation is trivial in this time-independent case, so we must use both the space-space and time-time components of the Einstein equations.  

At short distances the quadratic $\pi$ action (\ref{effectiveS}) is
\be
{\cal L}_\pi \simeq {M_4^2} \left[ 3m^2  \, \pi \Box \pi + 2m \, \pi (K_{\mu\nu} - g_{\mu\nu} K) \nabla^\mu \nabla ^\nu \pi  \right] \; ,
\ee
so the new kinetic matrix is proportional to
\be
3 m^2  \, g_{\mu\nu} + 2m (K_{\mu\nu} - g_{\mu\nu} K) = 
3  m^2 \, {\rm  diag} \left(- \left[ 1-\sfrac{2f}{mr}\right] f^2 ;   \left[ 1-\sfrac{2}{3mr}(f+1/f) \right] g_{ij} \right).
\ee
If $\pi$ is to have positive kinetic and gradient energies, we want both terms in brackets to be positive. For $\pi$ to be  subluminal, we want the second bracket to be smaller than the first. Indeed, the speed of $\pi$ excitations as measured in a local Minkowskian frame at rest with respect to the brane  is
\be
c^2_\pi = \frac{1-\sfrac{2}{3mr}(f+1/f)}{1-\sfrac{2f}{mr}} \; ,
\ee
as can be seen by locally going to Riemann normal coordinates $\hat x^\mu$,
\be
d \hat t^2 = f^2 dt^2 \; , \qquad d {\hat {\vec x}}^2 = g_{ij} dx^i dx^j = r^2 d \Omega_3^2 \; .
\ee

Specialize now to the case of a pure cosmological constant on the brane, $p=-\rho$.  From equating (\ref{E00}) and (\ref{Eii}) we find 
\be {2\over (mr)^2}+{1\over mr}\left({1\over f}-2f\right)=0.\ee
Given any value $m r>0$, this determines $f$, thus $m^2 \mu$.  The cosmological constant ${\rho\over M_4^2m^2}$ is then determined through either  (\ref{E00}) or (\ref{Eii}).   Thus, there is a single parameter family of possible static solutions, one for each $r>0$.  

The 5D mass $\mu$ is positive only for $mr > 2$. We want $\mu$ to be positive, otherwise the solution's 5D mass would be negative, thus signaling an instability against accelerating away. The cosmological constant is always negative, ranging monotonically from $-3/2 \, m^2 M_4^2$ to $0$ as $r$ ranges from $0$ to $\infty$.  Notice that the brane always lies outside the would-be Schwarzschild horizon. 

\begin{figure}[h!]
\begin{center}
\includegraphics[width=12cm]{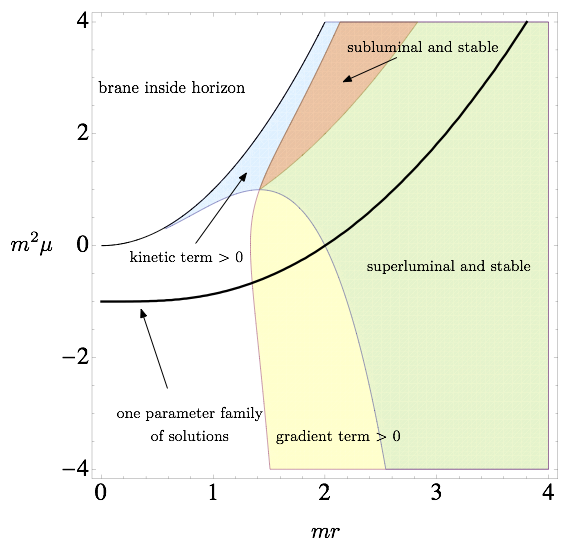}
\caption{\it \small Stability and superluminality of static solutions.  The thick black curve is the one-parameter family of solutions discussed in the text. 
The upper-left white region is where the brane is inside the Schwarzschild horizon---and of course there is no static solution there.}
\label{staticepsilon1}
\end{center}
\end{figure}

Figure \ref{staticepsilon1} shows the stability and superluminality of the solutions.  The $\pi$ mode has ghostlike (negative kinetic energy) and/or tachyonic (negative gradient energy) instabilities for $mr<2$, corresponding to $\mu<0$.  It becomes stable thereafter, but is always superluminal.   This means that in DGP, for values of the brane cosmological constant in the range $ -3/2 \, m^2 M_4^2 < \rho < 0$, there exists a solution which can not be thrown out on stability grounds, which nevertheless supports superluminal propagation.  As $r\to\infty$, the kinetic and gradient coefficients approach $1$ from below, and the speed of $\pi$ approaches $1$ from above.  Figure 2 plots the 4D energy density $\rho$ vs. the brane position $r$.

\begin{figure}[h!]
\begin{center}
\epsfig{file=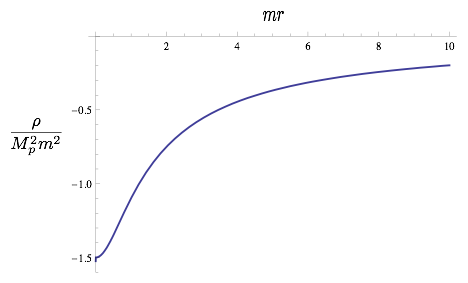, width=4.5in}
\caption{\it \small $\rho$ as a function of $r$ for static solutions.}
\label{staticepsilon-1}
\end{center}
\end{figure}

\subsection{Time dependent spherical solutions}\label{time_dep}
\ \ \ \ \
By studying static spherical solutions, we have seen that for a finite range of negative values of the brane cosmological constant, $ -3/2 \, m^2 M_4^2 < \rho < 0$, there exist stable superluminal solutions.  To obtain information about other values of the brane cosmological constant, we now search for brane solutions that are still spherical, but with a time dependent radius $r=a(t)$. For a 4D observer, these solutions describe a cosmology with positive spatial curvature. These backgrounds have been studied with somewhat different methods first  in \cite{deffayet} and subsequently in \cite{gkmp} .  Here we will study the stability and speed of propagation of their fluctuations.  

To start, take coordinates $t,\theta_i$ as brane coordinates.  $\theta_i$, $i=1\ldots 3$, are the angles for the three-sphere. The embedding is then
\be t=t, \ \ r=a(t),\ \ \theta_i=\theta_i.\ee
The induced metric is
\be ds^2=-\left[f^2-{1\over f^2}\left(da\over dt\right)^2\right]dt^2+a(t)^2d\Omega_{3}^2.\ee
Note that the term in brackets in front of $dt^2$ should be positive, to ensure that the brane is time-like.  This requires $\left|{da\over dt}\right|<f^2$, i.e. that the brane not be expanding or contracting as fast or faster than the speed of light in the bulk.   

The tangent vectors to the brane are $e^{A}_{\ \mu}={\partial X^A\over\partial y^\mu}$, where $X^A$ are the 5D embedding coordinates, and $y^a$ are the brane coordiantes.  We have,
\bea e^A_{\ t}&=&(1,{da\over dt},0,0,0),\\
e^A_{\ \theta_i}&=&(0,0,\hat{\mathbf{e}}_i).\\
\eea
The normal vector $n^A$ is orthogonal to all of these, is spacelike, and is normalized, i.e. $g_{AB} \, e^{A}_{\ \mu}n^B=0$, $g_{AB} \, n^An^B=1$.  
\be n^{A}={1\over f\sqrt{-\left({da\over dt}\right)^2+f^4}}\left({da\over dt}, f^4,0,0,0\right).\ee
Again, the bulk is taken to be outside. 

The extrinsic curvature can be calculated using $K_{\mu\nu}=e^A_{\ \mu}e^B_{\ \nu}\nabla_An_{B}=e^A_{\ \mu}\partial_\nu n_A-e^A_{\ \mu}e^B_{\ \nu}\Gamma^C_{AB}n_C$, 
\be K_{\mu\nu}=-{1\over \sqrt{-\left({da\over dt}\right)^2+f^4}}\left(f^4 f'-3f'\left({da\over dt}\right)^2+f{d^2a\over dt^2}\right)dt^2+{af^3\over \sqrt{-\left({da\over dt}\right)^2+f^4}}d\Omega_{(3)}^2.\ee
Here $f'={df\over da}$.

We want to bring the induced metric into FRW form.  To do this, define a new time coordinate $\tau(t)$ on the brane by solving
\be \label{coordtime} 
\left({d\tau\over dt}\right)^2=f^2-{1\over f^2}\left(da\over dt\right)^2
\qquad \Rightarrow \qquad
{dt\over d\tau}={1\over f^2}\sqrt{f^2+\left(da\over d\tau\right)^2} \; .
\ee
The embedding is now
\be 
t=t(\tau) \, \qquad r=a(\tau) \; ,\qquad \theta_i=\theta_i \; ,
\ee
and the induced metric is
\be 
ds^2=-d\tau^2+a(\tau)^2d\Omega_{3}^2 \; .
\ee
Some quantities we will need are  
\bea 
G_{00}&=&3\left({\dot a^2\over a^2}+{1\over a^2}\right) \; , 
\\ G_{ij}&=&-\left(1+\dot a^2+2 a\ddot a\right)\gamma_{ij} \; , \\ 
R_{ij}-{1\over 6}R \, g_{ij}&=&\left(1+\dot a^2\right)\gamma_{ij} \; .
\eea
Dot means ${d\over d\tau}$, and $\gamma_{ij}$ is the metric on the unit three-sphere.  The spatial part of the extrinsic curvature, in these coordinates, reads
\be 
K_{ij}= a\sqrt{f^2+\dot a^2} \gamma_{ij} \; .
\ee
(There's actually no need to calculate $K_{00}$, we can get away without it, as we'll see.)

Trace reversing (\ref{junctioncondition}), we have
\be \label{tracereversed} M_4^2\left(R_{\mu\nu}-{1\over 6}Rg_{\mu\nu}\right)-2M_5^3K_{\mu\nu}=T_{\mu\nu}-{1\over 3}Tg_{\mu\nu}.\ee
Using the quantities above, the spatial components give us the Friedmann equation,
\be \label{friedmann}
\left({\dot a^2\over a^2}+{1\over a^2}\right)-{m\over a}\sqrt{f^2+\dot a^2}={1\over 3M_4^2} \rho \; ,
\ee
where 
\be m={2M_5^3\over M_4^2}\ee
is the DGP scale and $\rho$ the total 4D energy density.  We also have the usual conservation equation
\be 
\dot \rho+3{\dot a\over a}\left(\rho+p\right)=0 \; ,
\ee
which together with (\ref{friedmann}) determines 
the $00$-component of (\ref{tracereversed}).  (Note that in the static case, the conservation equation is trivial, so there we did need the $00$-component of (\ref{tracereversed}) as well.)

We can find the on-shell value of $K_{00}$ by looking at the trace of (\ref{junctioncondition}),
\be 
-M_4^2 R+6 M_5^3 K=T= - \rho+3 p \; ,
\ee
which implies
\bea  
K & = &{1\over m}\left[{1\over 3M_4^2} \left(- \rho+3p \right)+2\left({\dot a^2\over a^2}+{1\over a^2}+{\ddot a\over a}\right)\right], \\
K_{00} & = & {3 \over a}\sqrt{f^2+\dot a^2}-K \; .
\eea
The coefficient of the $\pi$ kinetic term reads 
\be \label{scalarkinetic} 3m^2g_{\mu\nu}+2m(K_{\mu\nu}-g_{\mu\nu}K).\ee
The $00$-component is 
\be -3m^2\left[1-{2\over ma}\sqrt{f^2+\dot a^2}\right].\ee
The bracket must be positive if there are to be no ghosts. 
The $ij$-components are
\be 3m^2 a^2\gamma_{ij}\left[1-{2\over3ma}\sqrt{f^2+\dot a^2}-{2\over 3m^2}\left({\rho+3p\over 3M_4^2}+2{\ddot a\over a}\right)\right].\ee
The bracket must be positive if there are to be no tachyons.  
The local speed of $\pi$ is
\be c_\pi^2= {1-{2\over3ma}\sqrt{f^2+\dot a^2}-{2\over 3m^2}\left({\rho+3p\over 3M_4^2}+2{\ddot a\over a}\right)\over 1-{2\over ma}\sqrt{f^2+\dot a^2}}.\ee

Specialize now to the case of a pure cosmological constant on the brane, $-p=\rho$.  Differentiating the Friedmann equation (\ref{friedmann}), we find an equation which can be solved for $\ddot a$ in terms of lower derivatives.  There is a three parameter family of solutions; we can pick arbitrary initial data $a(0)$ and $\dot a(0)$, and some value for $\mu$.  The cosmological constant is then determined by (\ref{friedmann}).  The time derivative of (\ref{friedmann}) is used to evolve these initial data, creating a trajectory in the space of independent initial data spanned by $a,\ \dot a$ and $\mu$.   Note that the initial data must also satisfy $1-{\mu\over a^2}+\dot a^2\geq 0$ because of the square root in (\ref{friedmann}). 

\begin{figure}[b!]
\begin{center}
\includegraphics[width=10cm]{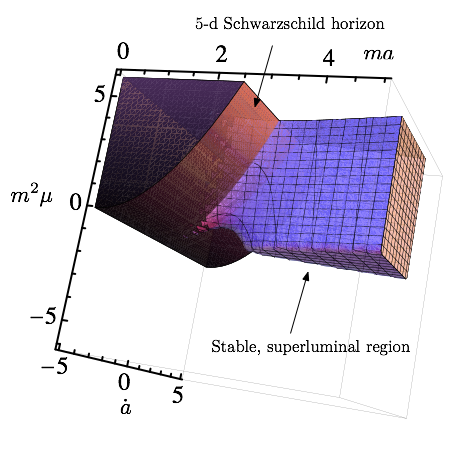}
\caption{\it \small $a,\ \dot a, \ \mu$ parameter space.}
\label{fullregionepsilon}
\end{center}
\end{figure}

\begin{figure}[b!]
\begin{center}
\includegraphics[width=13cm]{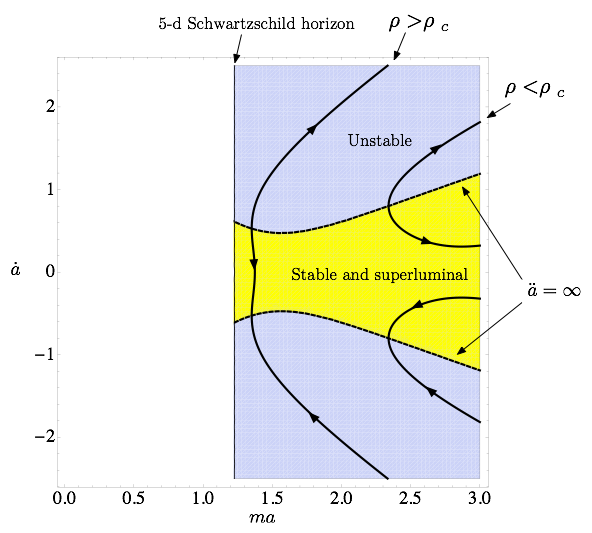}
\caption{\it \small Examples of stable, superluminal solutions.  Shown is a typical $\mu={const.}>0$ plane.  The solution is confined to move along a $\rho={const.}$ contour, shown here as solid black lines for two different values of $\rho$, with arrows indicating the direction of flow.  The solutions come to a 4D curvature singularity when $\ddot a = \infty$, which is exactly the boundary of the region in which $\pi$ is stable and superluminal.  There is a critical value $\rho_c={3M_4^2\over 2\mu}\left(1-\sqrt{1+m^2\mu}\right)$, at which the $\rho=const.$ line bifurcates into two pieces.  The leftmost line shown is for a value $\rho>\rho_c$, and the stable superluminal solution lives for only a finite proper time, going from singularity to singularity as it traverses the stable superluminal region.   The rightmost line is at a value $\rho<\rho_c$, and the stable superluminal solution lives for an infinite time, extending to infinite radius.  The values used for this plot are, in units of $m$, $\mu=1.5,\ \rho=\pm 0.6M_4^2$.}
\label{stablesuperluminal}
\end{center}
\end{figure}

Figure \ref{fullregionepsilon} shows the $a,\ \dot a, \ \mu$ parameter space.   The plot shows the region where $\pi$ is stable and superluminal.  We will try to rule out more values of the brane cosmological constant by asking whether they support solutions which stay in this region.   In addition, the solutions must be within the effective field theory where $\pi$ non-linearities can be ignored\footnote{Incidentally, the region in which $\pi$ is both stable and subluminal does not intersect the region $\mu>0$.  This means that in this class of solutions there are none which, at any time in their history, are subluminal, stable, and have positive energy from the 5D point of view ($\mu> 0$).}.  

We seek solutions with positive 5D energy so we restrict our attention to the region $\mu>0$.  The value of $\mu$ is fixed and conserved, so all the dynamics occurs within a $\mu={const.}$ plane.  This plane is shown in figure \ref{stablesuperluminal}.  Also shown is the region where $\pi$ is stable and superluminal.  We seek solutions that stay in this region for all times.  

The Friedmann equation (\ref{friedmann}) gives another surface with a fixed value of $\rho$ on which the dynamics is confined.  The intersection of these surface with the $\mu={const.}$ plane then determines the dynamical path.  The time parametrization along the path is determined through the evolution equation given by differentiating the Friedmann equation with respect to time and solving for $\ddot a$.  The evolution can in fact come to an end at points where $\ddot a=\infty$.  This can happen at points where $a,\dot a$ are finite, so this represents a 4D curvature singularity, as can be seen from the Ricci scalar $R=6\left({\ddot a\over a}+{\dot a^2\over a^2}+{1\over a^2}\right)$ (these are the pressure singularities discussed in \cite{gkmp}).  As it turns out, the line on which $\ddot a$ blows up is exactly the border of the stable superluminal region.  This means that solutions can never cross into or out of this region. 

For any value of $\rho$ there are solutions which live out their lives in the stable superluminal region.  For $\rho>{3M_4^2\over 2\mu}\left(1-\sqrt{1+m^2\mu}\right)$, they come into existence out of a pressure singularity  at one border of the stable superluminal region, travel through the region, and then come to an end in a pressure singularity on the other side of the region.  For $\rho<{3M_4^2\over 2\mu}\left(1-\sqrt{1+m^2\mu}\right)$, the solutions have a pressure singularity in either the past or the future, and live for an infinite proper time, either coming in from or going out to $a=\infty$.  
Thus, for \textit{any} value of $\rho$, there exist solutions which cannot be thrown out on stability grounds, and yet have superluminal propagation.  

As a consistency check of our results, we want to recover the celebrated `self-accelerating' solution---a zero-tension de Sitter brane embedded in a Minkowski bulk.
Therefore, we look for an empty brane ($\rho=0$) in a flat bulk ($\mu=0$).   The Friedmann equation (\ref{friedmann}) reduces to 
\be 1+\dot a^2= m a \, \sqrt{1+\dot a^2}.\ee
Integrating the Friedmann equation yields the unique (up to time translations) solution
\be a(\tau)={1\over m}\cosh (m\tau).\ee
Rewriting in terms of the coordinate time $t$, using (\ref{coordtime}), we obtain
\be a(t)={1\over m}\sqrt{1+m^2t^2}.\ee
This is the familiar self-accelerating solution, a hyperboloid in a flat bulk, where the bulk is taken to be the outside.  
The hyperboloid is maximally symmetric, embedded in flat Minkowski space, so its extrinsic curvature is proportional to the induced metric,
\be K_{\mu\nu}=mg_{\mu\nu},\ee
and the scalar kinetic matrix (\ref{scalarkinetic}) becomes
\be 3m^2g_{\mu\nu}+2m(K_{\mu\nu}-g_{\mu\nu}K)=-3m^2g_{\mu\nu}.\ee
Thus the extrinsic curvature corrections have the effect of \emph{reversing} the sign of the kinetic term, recovering the well known fact that $\pi$ is a ghost, propagating at exactly the speed of light \cite{LPR}.

\section{The retarded Green's function}\label{green}
\ \ \ \ \
We now come back to the issue of non-local corrections to the 4D effective action (\ref{effectiveS}). In particular, we want to determine the support of $\pi$'s retarded boundary-to-boundary Green's function. This is enough to study the causal response of $\pi$ to local boundary sources, i.e.~to study the causality properties of the interaction between sources mediated by $\pi$. We will see that to this purpose, the conclusions based on the local part of the action are correct.

To simplify the analysis,  consider a DGP-like theory for a scalar, with a localized superluminal ($c>1$) kinetic term on the boundary:
\be \label{phi_eq}
\Box_5 \pi (x,y) + \delta(y) \, \tilde \Box_4 \pi(x,y) = J(x) \delta(y) \; ,
\ee
where
\be
\tilde \Box_4 \equiv -\di_t^2 + c^2 \nabla^2 \
\ee
and $J(x)$ is the boundary source. For simplicity we are suppressing the dimensionful constant $m$ that sets the critical scale. So, here the critical scale is one. This example is much simpler than DGP, but it captures the features that are relevant for our discussion.

Taking the Fourier transform in 4D, the Green's function
\be
G(x,y) = \int \frac{d^4p}{(2\pi)^4} \, e^{+i p \cdot x} \, \hat G(p, y)
\ee
obeys the equation
\be
\big[ (p^2 - \di_y^2) +  \delta(y) \tilde p ^2 \big] \hat G(p, y) = \delta(y) \; ,
\ee
with 
\be
\tilde p^2 \equiv -\omega^2 + c^2 \vec p \, ^2 \; .
\ee
The solution that vanishes at $y \to \pm \infty$ for Euclidean four-momenta, $\vec p \,^2 > \omega^2$, is
\be \label{greens}
\hat G(p, y) = - \frac{e^{-\sqrt{\vec p^2 - \omega^2} |y|}}{-\omega^2 + c^2 \vec p \, ^2 + 2 \sqrt{\vec p \,^2 - \omega^2}}  \; .
\ee
In the complex $\omega$ plane, for real $\vec p$ this function has branch cuts starting at $\omega = \pm |\vec p|$.
We define the square root in such a way that  the branch cuts run along the real axis from the branch points to $\omega = \pm \infty$, respectively. This corresponds to setting the branch cut of $\sqrt{z}$ along the negative real $z$-axis:
\be
\sqrt{\rho \, e^{i \alpha}} \equiv \sqrt{\rho} \, e^{i \alpha/2} \; , \qquad \mbox{for} \quad -\pi < \alpha < \pi
\ee 

$\hat G(p, y)$ has poles too, for
\be \label{poles}
-\omega^2 + c^2 \vec p  \,^2 + 2 \sqrt{\vec p \,^2 - \omega^2} =0  \; .
\ee
However it is easy to convince oneself that in the superluminal case $c^2 > 1$ these lie on the second sheet---i.e., they correspond to zeroes of the denominator in (\ref{greens}) for the other branch of the square-root.
To see this, define
\be
z \equiv \sqrt{\vec p \,^2 - \omega^2} \; .
\ee
Given our definition of the square root, $z$ takes values in the right half-plane, ${\rm Re} (z) \ge 0$. Now, the pole condition (\ref{poles}) reads
\be
z^2 + 2 z + (c^2-1) \vec p \,^2 = 0 \; ,
\ee
which for real $\vec p$ and $c^2$ larger than one, has no solutions in the right half-plane.

The fact that there are no poles in the physical sheet of $\omega$ tells us that there are no {\em free} solutions for $\pi$ with the given boundary conditions at $|y| \to \infty$. This is hardly surprising, for we already know that in DGP there is no mode localized on the boundary. Rather, four-dimensional excitations behave like resonances, ``leaking'' into the bulk with a timescale of order $m^{-1}$. Nevertheless, the Green's function for $\pi$ is all we need to  compute the interaction between local sources that couple  to it, regardless of whether $\pi$ describes an asymptotic state of the theory.

The Green's function in real space is
\be
G(x,y) =   -  \int \frac{d^4p}{(2\pi)^4} \, \frac{e^{-i \omega t + i \vec p \cdot \vec x -\sqrt{\vec p \,^2 - \omega^2} |y|}}{-\omega^2 + c^2 \vec p  \,^2 + 2 \sqrt{\vec p \,^2 - \omega^2}}  
\ee
\begin{figure}[t!]
\begin{center}
{\small ({\it a}) \hspace{8cm} ({\it b})} \\[.2cm]
\includegraphics[angle=-90,width=7.5cm]{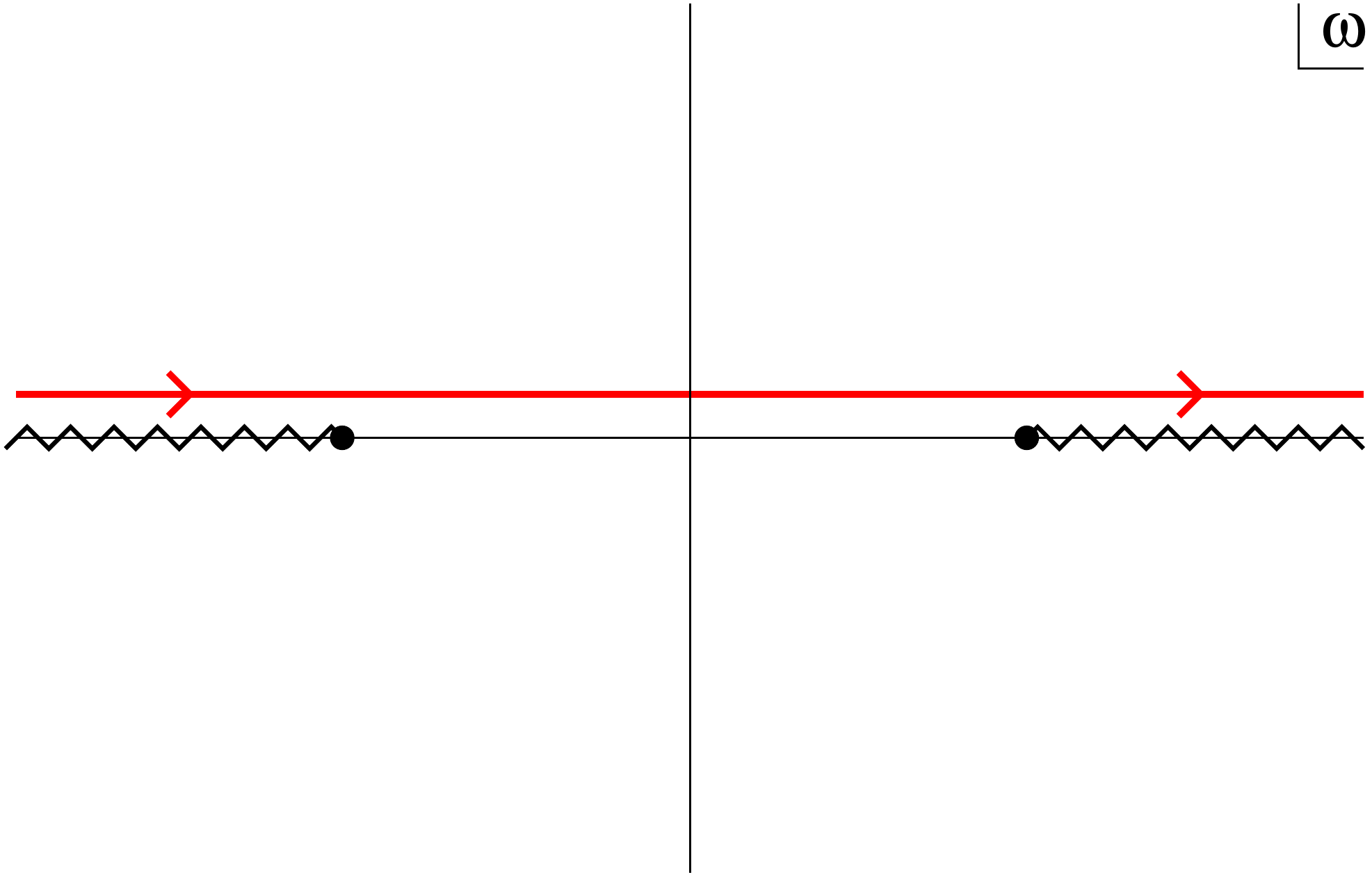}
\hspace{1cm}
\includegraphics[angle=-90,width=7.5cm]{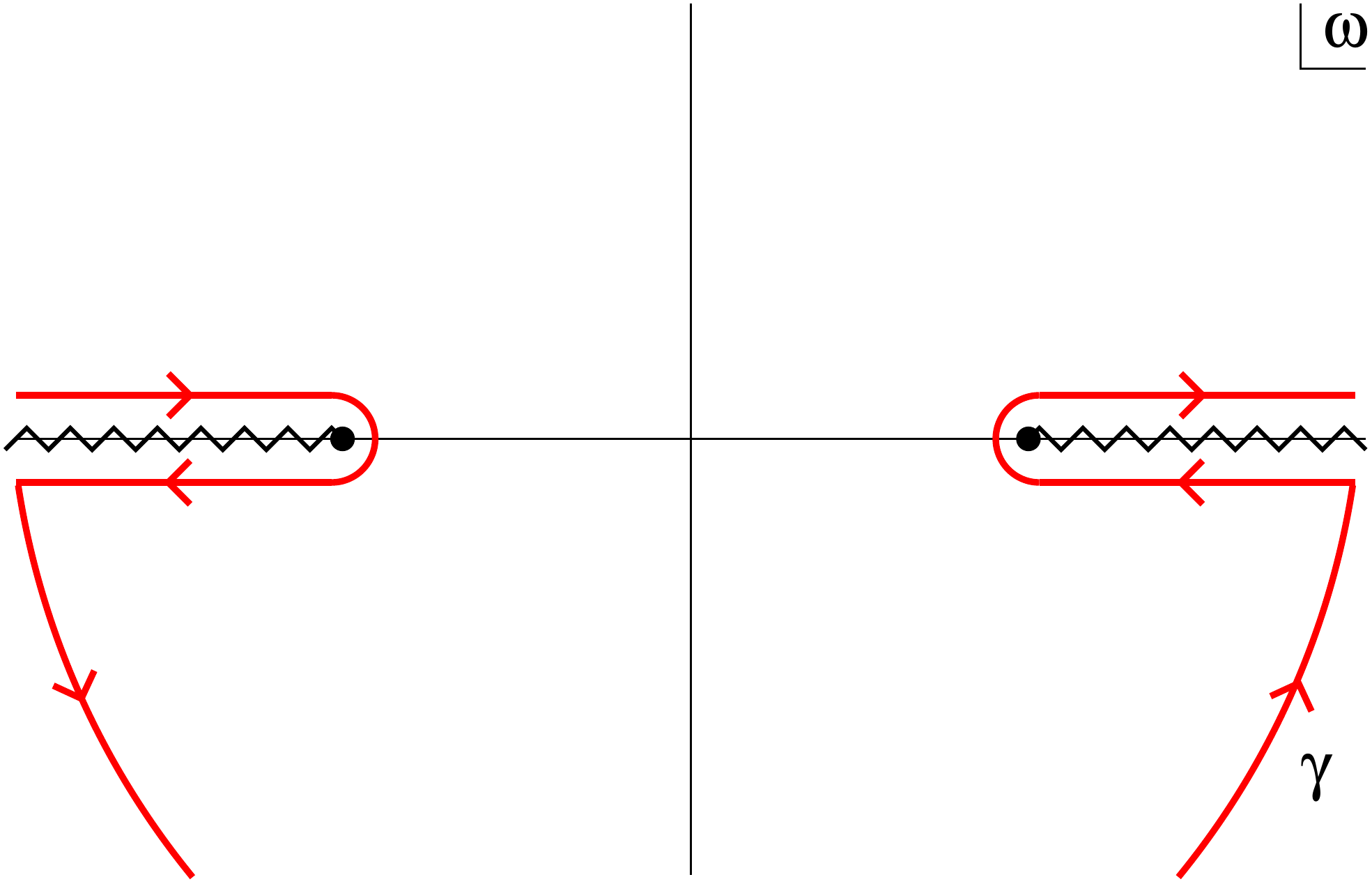}
\caption{\em \small The integration contours for the integral in $\omega$, as  described in the text. The branch points are at $\omega = \pm k \equiv \pm |\vec p|$. \label{omega_integrals}}
\end{center}
\end{figure}
To define the retarded Green's function, we impose the condition that $G$ vanish for $t < 0$. If we integrate just above the cuts on the real $\omega$ axis, as shown in fig.~\ref{omega_integrals}{\it a}, we can close the contour at infinity in the upper half-plane, where we have no singularity. At infinity the integrand behaves like
\be \label{exponentials}
e^{-i \omega t} e^{ - z |y|} \; ,
\ee
where, as we discussed above, $z$ has a positive real part. This means that $G(x,y)$ vanishes for $t < 0$, as desired. In conclusion, the retarded Green's function corresponds to integrating above the cuts in the $\omega$ plane.

We want to see what happens for $t>0$. First, notice that the retarded Green's function vanishes in the bulk for $|y| > t$. This has to be expected, for we do not anticipate any superluminal effect in the direction orthogonal to the boundary.
To see that this is indeed the case, suppose we still compute the $\omega$-integral by closing the contour in the upper half-plane. The integrand's behavior is still given by (\ref{exponentials}), with $t > 0$. For fixed real $\vec p$ and large $\omega$ in the upper half plane we have
\bea
|\omega| \gg |\vec p| \; , \quad {\rm Im}(\omega) > 0    \quad & \Rightarrow &  z \to -i \omega \sqrt{1 - \vec p \, ^2/ \omega^2} \\
&& e^{-i \omega t} e^{ - z |y|} \sim e^{ i \omega (|y| - t)} 
\eea
which decays as long as $|y| > t$, and we get
\be
G_{\rm ret}(x,  |y|>t) = 0 \; ,
\ee
as expected.

Notice that in the case of $\pi$, what happens in the bulk is not physically relevant, since $\pi$ is pure gauge there. More interesting, and more complicated, is the propagation parallel to the boundary. We focus on the  $y = 0$ case, which 
corresponds to the boundary-to-boundary propagator.
First, we can integrate over the angular variables of $\vec p$. Defining $k \equiv |\vec p|$ and $r = |\vec x|$ we get
\be \label{greens2}
G_{\rm ret}(x,y = 0) =   - \frac1{2 \pi^3} \frac1r \int_0^\infty \! dk \,  k \sin (kr) \int \! d \omega \, \frac{e^{-i \omega t} }{-\omega^2 + c^2 k^2 + 2 \sqrt{k^2 - \omega^2}}  \; .
\ee
Then, for $t > 0$, we can integrate in $\omega$ by deforming the contour in the lower half-plane as shown in fig.~\ref{omega_integrals}{\it b}. The contribution from infinity vanishes, and we are left with the integral along the cuts:
\bea 
G_{\rm ret}(x,y=0) & = &   - \frac1{2 \pi^3} \frac1r \int_0^\infty \! dk \,  k \sin (kr)  \times  \nonumber \\ 
&& \bigg( \int_{-\infty}^{-k} \! d \omega \, + \int^{\infty} _k \! d \omega \bigg)
e^{- i \omega t} \Delta \bigg[    \frac{1}{-\omega^2 + c^2 k^2 + 2 \sqrt{k^2 - \omega^2}} \bigg] \; , \label{Delta}
\eea
where $\Delta[\cdots]$ denotes the discontinuity across the cut---the value above minus the value below.
For our definition of the square root, we have
\bea
k < \omega < \infty & \Rightarrow & \sqrt{k^2 - \omega^2} \big |_{\rm above}  =   - i \zeta  \label{positive}\\
&& \sqrt{k^2 - \omega^2} \big |_{\rm below}  =  + i \zeta \; , \qquad \mbox{with} \; \zeta \in \mathbb{R}^+ \\
-\infty < \omega < -k & \Rightarrow & \sqrt{k^2 - \omega^2} \big |_{\rm above}  =   - i \zeta \\
&& \sqrt{k^2 - \omega^2} \big |_{\rm below}  =  + i \zeta \; , \qquad \mbox{with} \; \zeta \in \mathbb{R}^-\; .
\eea
Also
\be
\zeta^2 = \omega^2 - k^2 \quad \Rightarrow \quad \zeta \,  d \zeta = \omega \,  d \omega  \; .
\ee
The two $\omega$-integrals in (\ref{Delta}) thus combine to yield an integral all along the real $\zeta$ axis,
\bea
G_{\rm ret}(x,y=0)  & = &   - \frac1{2 \pi^3} \frac1r \int_0^\infty \! dk \,  k \sin (kr) \int_{-\infty} ^\infty \! d \zeta \, \frac{e^{-i \omega t}}{\omega} \frac{4i \, \zeta^2}{\big[ \zeta^2 - (c^2-1)k^2 \big]^2 + 4 \zeta^2} \nonumber \\
= & \!\! - & \!\!\!\!\!  \frac2{ \pi^3} \frac1r  \int_0^\infty \! dk \,  k \sin (kr) \int_{-\infty} ^\infty \! d \zeta \, \frac{\sin (\omega t)}{\omega}
\frac{ \zeta^2}{\big[ \zeta^2 - (c^2-1)k^2 \big]^2 + 4 \zeta^2}  
\eea
where for real $\zeta$
\be
\omega = {\rm sign}( \zeta) \sqrt{\zeta^2 + k^2} \; ,
\ee
and in the second step we only kept the even part of the integrand under $\zeta \to -\zeta$. What is interesting about this form of $G_{\rm ret}$ is that $\sin (\omega t)/ \omega$, as an even function of its argument, is analytic in the whole $\zeta$ and $k$ planes, and we do not have to worry about the cuts of $\omega$ as a function of $\zeta$ and of $k$. 

\begin{figure}[b!]
\begin{center}
{\small ({\it a}) \hspace{8cm} ({\it b})} \\[.2cm]
\includegraphics[angle=-90,width=7.5cm]{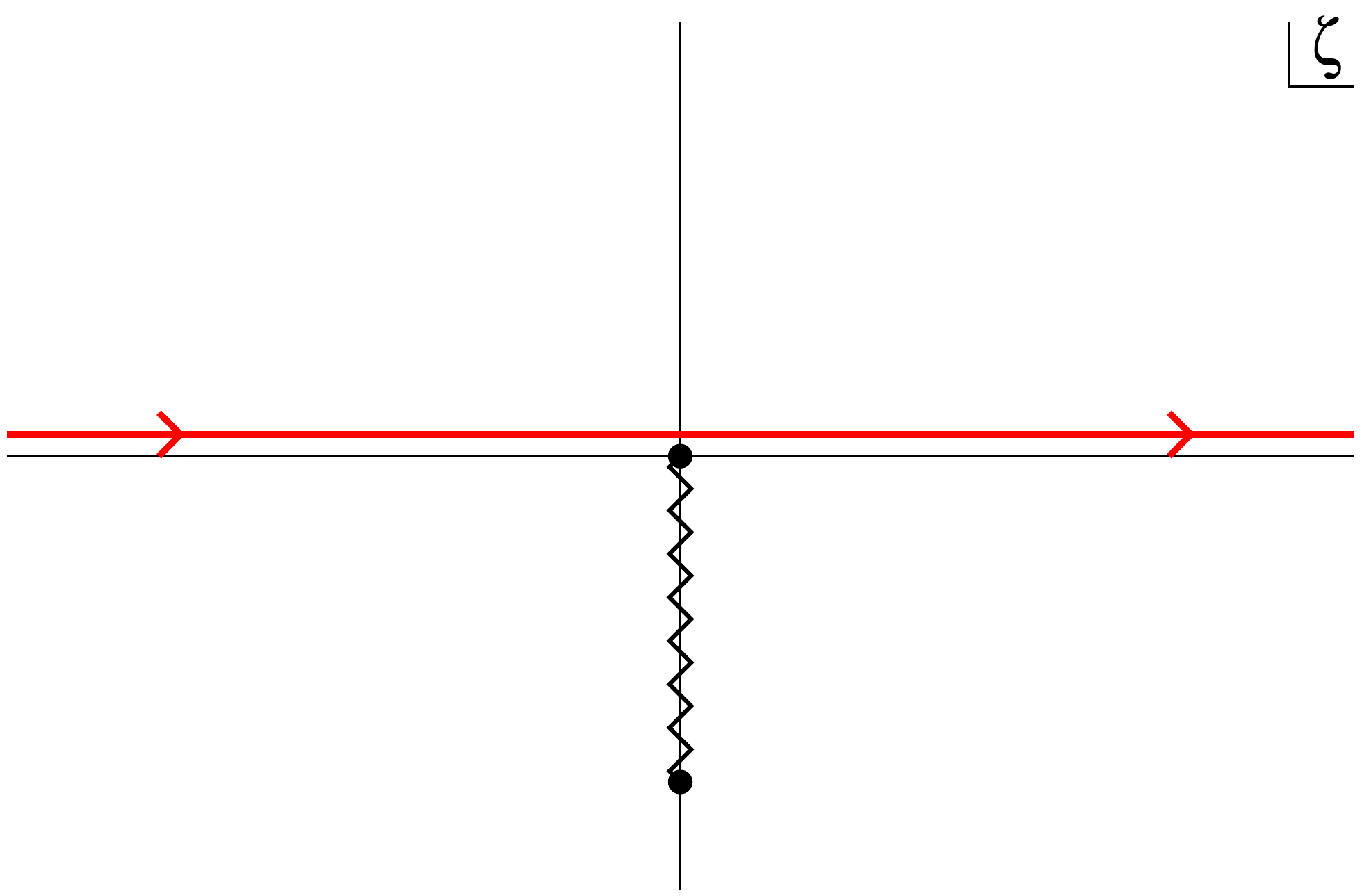}
\hspace{1cm}
\includegraphics[angle=-90,width=7.5cm]{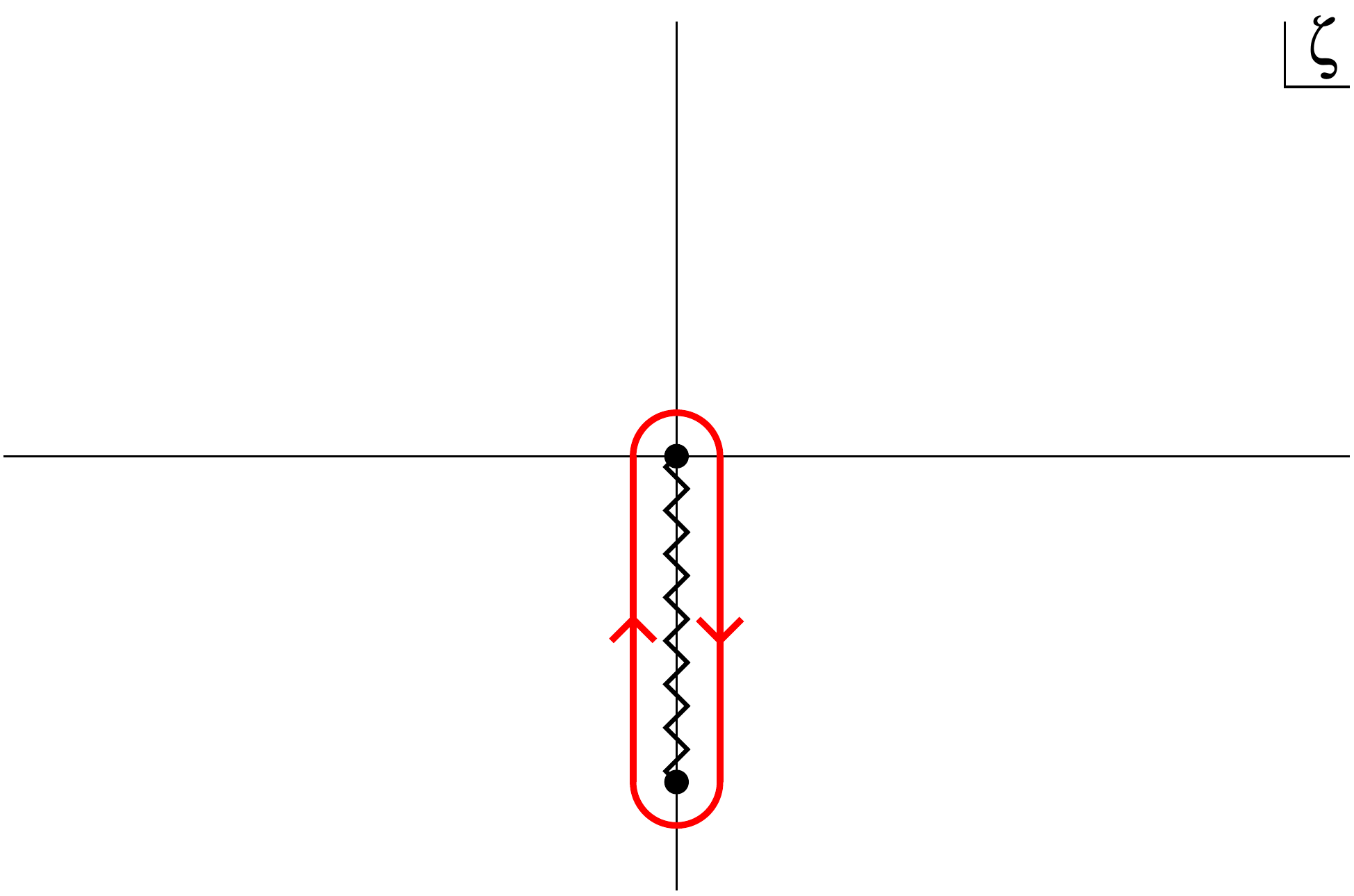}
\caption{\em \small The integration contours for the integral in $\zeta$, as  described in the text. The branch points are at $\zeta = 0 $ and $\zeta = -2i$. \label{zeta_integrals}}
\end{center}
\end{figure}

In the $k$ plane, we have four poles for 
\be
k^2 = \frac{\zeta^2 \pm 2i \zeta}{c^2-1} \; .
\ee
However, before integrating in $k$ we have to extend the range of integration from $-\infty$ to $+\infty$, which given that the whole integrand is even in $k$, is straightforward. Then, in $\sin(kr)$ we can keep just one of the two exponentials---for instance we can perform the replacement $\sin(kr) \to \sin(kr) - i \cos(kr) = - i e^{i kr}$, which is allowed because of parity.
We get
\be
G_{\rm ret}(x,y=0)  = \frac{i}{\pi^3} \frac1r \int_{-\infty} ^{\infty} \! dk  \int_{-\infty} ^{\infty} \! d \zeta
\, \frac{\sin (\omega t)}{\omega}
\frac{ k e^{ikr} \, \zeta^2}{\big[ \zeta^2 - (c^2-1)k^2 \big]^2 + 4 \zeta^2} \; .
\ee
Now, since we are interested in studying the propagation of superluminal signals, we can restrict to the $ r > t$ case. That is, we study the retarded Green's function outside the usual relativistic light-cone. In this case, we can close the contour in the upper half-plane of $k$. There, at large $k$ and fixed real $\zeta$ the integrand behaves as
\be
e^{i \, k r} e^{ \pm i \sqrt{\zeta^2 + k^2} \, t } \; ,
\ee
which decays for $r > t$. Therefore we get no contributions from infinity, but we do get two pole contributions, from
\bea
k_+  & =  & \zeta \sqrt{\frac{1 + 2 i /\zeta}{c^2 -1}}  \equiv \alpha(\zeta)  \\
k_-   & =  & - \zeta \sqrt{\frac{1 - 2 i /\zeta}{c^2 -1}}  =  \alpha(- \zeta) 
\eea
which for real $\zeta$ are both in the upper half-plane. Defining
\be
\beta  (\zeta) \equiv \zeta \sqrt\frac{ c^2 +2i / \zeta}{c^2-1}  \; ,
\ee
we have
\be \label{finalG}
G_{\rm ret}(r> t, y=0)  =  \frac{i}{4 (c^2-1) \pi^2} \frac1r \int_{-\infty} ^{\infty} \! d \zeta \bigg[ \zeta \, e^{ir \alpha(\zeta)} \,  \frac{ \sin ( t \, \beta(\zeta )) }{\beta(\zeta)}  + (\zeta \to -\zeta ) \bigg] \; .
\ee
Given the parity of the integrand in $\zeta$, we can keep either term in brackets, modulo an overall factor of two. We will keep the first.
As before, the factor $\sin(t \, \beta(\zeta))/\beta(\zeta)$ is analytic in the whole $\zeta$-plane. On the other hand, $\alpha(\zeta)$ has two branch-points,
\be
\zeta = 0 \; , \qquad \mbox{and} \quad \zeta = -2 i \; .
\ee
Our choice for the square root corresponds to making the branch-cut run between the two branch-points along the imaginary axis, as shown in fig.~\ref{zeta_integrals}{\em a}. 
%
%


If $r > ct$, we can close the contour in the upper half plane, where  the integrand is analytic. The exponential behavior at infinity is
\be
e^{i \, r \alpha(\zeta)} e^{\pm i \, t \beta(\zeta)} \simeq \exp \big( \sfrac{1}{\sqrt{c^2-1}} \, i \zeta \cdot (r \pm ct)\big)
\ee
which decays in the upper half plane as long as $r>ct$:
\be
G_{\rm ret}(r> c t, y=0) = 0 \;.
\ee
This shows that---as we expected---we can have superluminal signals traveling along the boundary, but they are confined to a new light-cone, with aperture determined by $c$.

As a final check that the system is well-behaved, we must make sure that {\em inside} this new light-cone the retarded Green's function is not exponentially increasing in time---which would signal an instability of the vacuum upon local perturbations. This is simple to check in the interspace between the two light-cones, i.e.~for $t < r < ct$. In this case we can still use (\ref{finalG}). We now have to split the sine into two exponentials,
\be
\frac{\sin (\beta t)} {\beta}= \frac{1}{2i \beta} \big( e^{i \beta t} - e^{-i \beta t} \big) \; ,
\ee
and analyze the two contributions separately. However, now for each individual contribution the analytic structure of $\beta(\zeta)$ matters, and we have to keep track of that as well. $\beta(\zeta)$ has a branch-cut running along the imaginary axis between $0$ and $-2i/c^2$. For $c^2 > 1$, this is entirely contained in the branch-cut of $\alpha(\zeta)$. Now, for the term $\sim  e^{ir \alpha(\zeta)} e^{it \beta(\zeta)}$  we can still close the contour in the upper half-plane. The behavior at infinity is
\be
\exp \big( \sfrac{1}{\sqrt{c^2-1}} \, i \zeta \cdot (r + ct)\big) \; ,
\ee
which vanishes for positive $r$ and $t$. We thus get zero from this term. For the other term we have to close the contour in the lower half-plane. We still get a vanishing contribution from infinity as long as $r < ct$---as we are assuming---but upon deforming the contour as in fig.~\ref{zeta_integrals}{\em b}, we are left with an integral along the cut:
\bea
G_{\rm ret}( t < r < ct,y=0) & = &   \frac{1}{4 (c^2-1) \pi^2} \frac1r \int_0 ^{-2i} \! d \zeta \, \zeta  \, \Delta   \bigg[ \frac{e^{i \alpha(\zeta) r - i \beta(\zeta) t}}{ \beta(\zeta)}   \bigg] \; ,  \\
&\propto& \int_0 ^{1} \! d w \, w  \, \Delta   
\bigg[ \frac{\exp \frac{w}{\sqrt{c^2-1}} \big( r \sqrt{1 -1/w}  -  t \sqrt{c^2 -1/w} \big)} { \sqrt{c^2 -1/w} }   \bigg] 
\eea
where in the second step we performed the change of variable $\zeta = - 2 i \, w$. The question now is whether this integral yields an exponential in time or $r$, with a positive real part in the exponent. 
We can split the integration region into two intervals, $0 < w < 1/c^2$ and $1/c^2 < w < 1$. In the former, both square roots in the exponent are purely imaginary, and we get an oscillatory contribution. In the latter interval the square-root that multiplies $r$ is still imaginary, but that multiplying $t$ is real. However, our definition for the square root has a non-negative real part in the whole complex plane. Therefore this contribution is oscillatory in $r$ and exponentially decaying in $t$.

In summary, we have shown that the retarded Green's function for $\pi$
\begin{itemize}
\item[{\em i)}]vanishes for $|y| > t$, i.e.~there is no superluminality in the direction orthogonal to the boundary;
\item[{\em ii)}] vanishes for $y=0$ and $r > c t$, i.e.~we have superluminality on the boundary, but we still have a causal cone with a finite aperture;
\item[{\em iii)}] is oscillatory inside this new causal cone---at least outside the usual relativistic light-cone---i.e.~the system is stable against local perturbations.
\end{itemize}
These are exactly the properties that one would have guessed from $\pi$'s e.o.m.~(\ref{phi_eq}).

\begin{figure}[t!]
\begin{center}
\epsfig{file=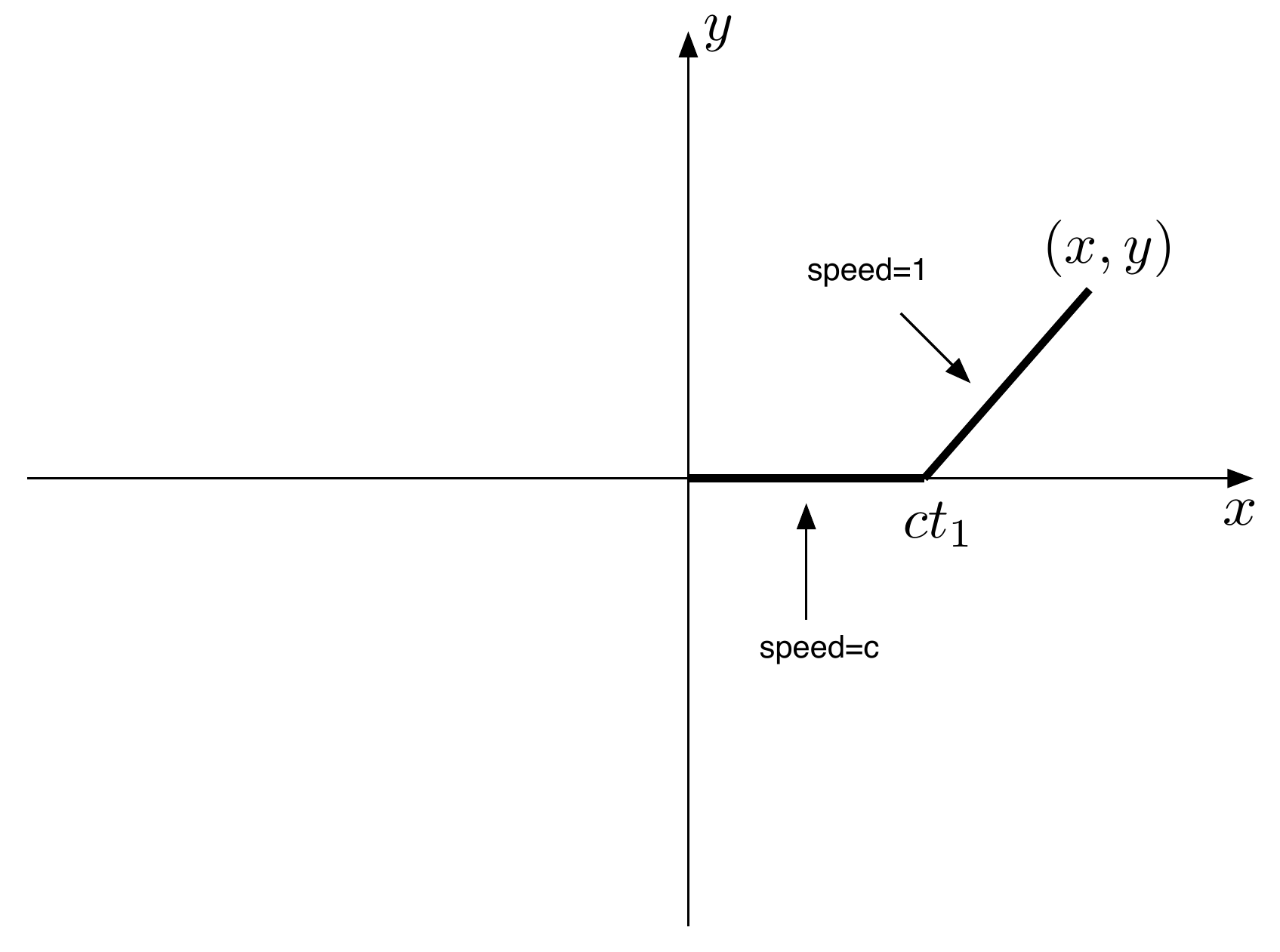, width=4in}
\caption{\it \small Light ray traveling from the origin.}
\label{lightrays}
\end{center}
\end{figure}

\subsection{The shape of the Green's function}
\ \ \ \ \
To conclude this section and generalize our results above to generic positions in the bulk, we wish to find the shape of the support of the retarded Green's function, with source centered at the origin. In the geometric optics approximation, its boundary at time $t$, as a function of $x$, will be the largest $y$ value a light ray starting at the origin can reach given time $t$.  We checked numerically that indeed the shape determined this way is correct. As usual, the causal structure is determined by the high-frequency limit.

Let a light ray start at the origin and travel along the brane for a time $t_1$ at the speed $c$, reaching the point $x=ct_1$.  It then leaves the brane and travels to point $(x,y)$ along a straight path at a speed of 1 (see figure \ref{lightrays}).  
To find the optimal time $t_1$ for the light ray to leave the brane, we wish to maximize $y$, given $x$ and fixed total travel time $t$, as we vary over $t_1$.  We have
\be y^2+(x-ct_1)^2=(t-t_1)^2,\ee
which has a maximum at 
\be t_1={cx-t\over c^2-1},\ \ \ y={ct-x\over \sqrt{c^2-1}}.\ee
Thus to reach the largest $y$ value, the light ray should travel along the brane for a time $t_1={cx-t\over c^2-1}$, and then leave the brane and make a bee-line for the point $(x,{ct-x\over \sqrt{c^2-1}})$.  We see that $t_1\rightarrow 0$ at $x={t\over c}$, so at this value of $x$, it becomes most favorable to leave the brane immediately.  

\begin{figure}[t!]
\begin{center}
\epsfig{file=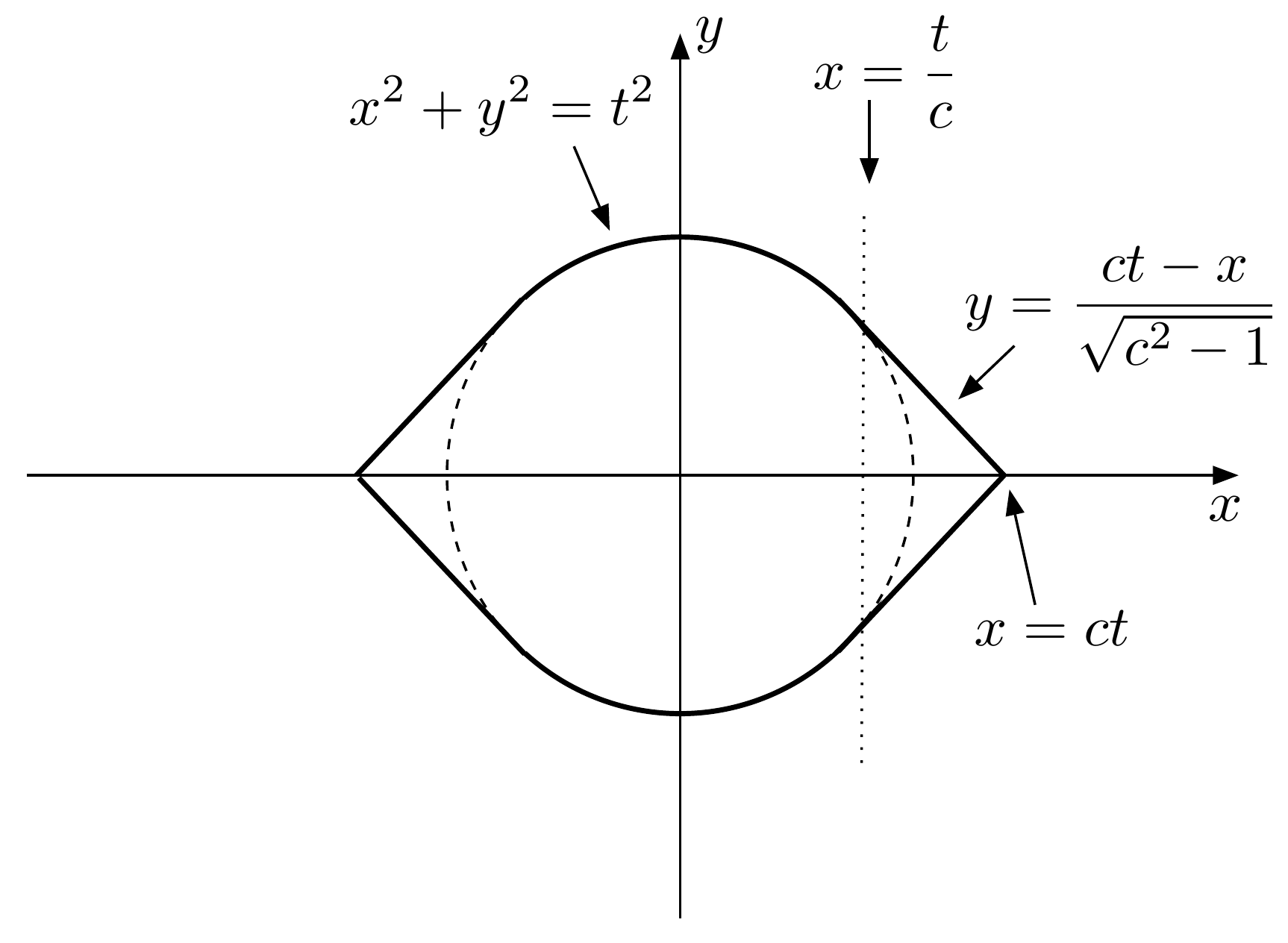,width=4.5in}
\caption{\it \small Shape of the support of the Green's function at time $t$.}
\label{greensfunction}
\end{center}
\end{figure}

The shape of the Green's function is shown in figure \ref{greensfunction}.  From $x={t\over c}$ to $x=ct$, where it is most favorable to travel down the brane a bit before leaving, it is the straight line $y={ct-x\over \sqrt{c^2-1}}$.  From $x=0$ to $x={t\over c}$, where it is most favorable to leave the brane immediately, it is just the circle, $y^2+x^2=t^2$.  The circle and the line intersect at $x={t\over c}$ and have the same slope there.  
This double teardrop shape represents the boundary of the support of the retarded Green's function.  It expands linearly in $t$.  It is of course rotationally symmetric among all of the $x$ coordinates.


\section{Discussion}
\ \ \ \ \
Superluminality on a non-Lorentz invariant background 
does not imply {\em per se} a breakdown of causality in
an effective field theory. It is nevertheless a troublesome signal which may 
indicate the absence of a relativistically invariant UV completion of the 
theory \cite{nima}. In this paper we showed that the DGP model exhibits superluminal
propagation in otherwise physically sound backgrounds, that are {\em exact}
solutions to the DGP equations. By computing small fluctuations of the scalar
mode around
exact solutions, we made sure that the pathology found here, which is of the
same nature as that found in~\cite{nima,nrt}, is  not an artifact
of the decoupling limit $M_4\rightarrow \infty$, $\Lambda_3\sim M_5^2/M_4=const$. 
The scalar mode couples locally to physical sources, namely 
to the trace of the stress-energy tensor, but its equations of motion are 
non-local away from the strict decoupling limit. 
To be sure that the maximum speed of
propagation we found using the local limit was a real effect, we computed the  
retarded Green's function of the scalar mode and found that on the 4D brane
it is non-vanishing in an enlarged light cone. We finally showed that within
this enlarged light cone, the Green's function is free of instabilities. This 
last check is important because it proves that the effect we found is truly
a maximum speed of propagation of signals that exceeds $c$, instead of being a 
tachyon-like instability of the background.

If we take the mildly conservative point of view that any theory that possesses
a relativistically-invariant, causal completion should not allow superluminality,
then we can use the results we found to rule out significant parts of the 
parameter space of DGP. Indeed, already the simple analysis 
presented in this paper shows that superluminality occurs whenever the
4D stress-energy tensor is a pure cosmological constant 
$T_{\mu\nu}=-\rho g_{\mu\nu}$, irrespective of
the value of $\rho$. Of course our analysis relies on the existence of a
domain of validity for the DGP model. DGP is after all an effective theory
which gets corrected by higher-dimensional operators at length scales
$L_5 \sim M_5^{-1}$.\footnote{\label{strong_coupling} Strictly speaking, there is a longer strong-coupling scale in the $\pi$ sector of the theory, $ \Lambda_3^{-1} \sim M_4/M_5^2 \gg L_5$ \cite{LPR}.
Nevertheless, classical solutions with (extrinsic) curvature in the strong-coupling regime are not necessarily outside the regime of validity of the effective theory \cite{nr}.
}
To carry on our analysis we need of course 
$L_{\rm DGP} \sim M^2_4 / M_5^3 \gg L_5$. So, one way to escape the results of this paper
is to have a 4D Planck mass $M_4$ not too much larger than
$M_5$. This is what one finds by trying to construct DGP as the
low-energy theory of a set of a few D-branes (or other kind of solitons) embedded
in a higher-dimensional string theory. 
On the other hand, a large hierarchy $M_4 \sim  N M_s \gg M_5 = M_s/g^{2/3}_s$ (with $M_s$ the string scale and $g_s$ the string coupling) might be 
achieved by considering a theory with ${\cal O}(N)$ D-branes when 
$N \gg 1/g_s^{2/3}$. Indeed, a hierarchy of scales between $M_4$ and the UV cutoff in a
theory with many species is a universal result that holds well beyond
the DGP context~\cite{drsv}. 
In this particular realization the  DGP scale is
\be
L_{\rm DGP} \sim \frac{M_4^2}{M_5^3} \sim \frac{N^2 g_s^2}{M_s} \; .
\ee
For our analysis to be valid we need $L_{\rm DGP} \gg 1/M_s$, that is 
\be
N g_s \gg 1 \; .
\ee 
In this case, the effective DGP action breaks down
whenever curvatures are of order of the string scale.
The tension of the D-brane arrangement---the four-dimensional cosmological constant energy density, $\rho$,---has a classical contribution of order $N M_s^4/g_s$, and a quantum-mechanical one of order $N^2 M_s^4$. However for the latter we may hope for cancellations due to supersymmetry, so let us ignore it. The intrinsic curvature of the D-brane system can be small in our limit
\be
R \sim \frac{\rho}{M_4^2} \sim \frac{M_s^2}{N g_s} \ll M_s^2 \; .
\ee
Still, the {\em extrinsic} curvature is always large,
\be
K_{\mu\nu} \sim \frac{\rho}{M_5^3} \sim N g_s \, M_s \gg M_s \; , \label{extrinsic}
\ee
thus signaling a breakdown of the effective field theory. In estimating the extrinsic curvature we used Israel's junction condition.  In principle, there could be a cancellation between the localized DGP term and the cosmological constant, but in general we do not expect such a cancellation of the leading contribution (\ref{extrinsic}).  Indeed, such a cancellation does not take place for the solutions discussed in sects.~\ref{static}, \ref{time_dep}, where the extrinsic curvature is always of order $\rho/M_5^3$. This is good, because those solutions are stable and have superluminal excitations---which we do not expect in a consistent D-brane arrangement in string-theory.



In conclusion, in this simple construction there is no choice for the values of $N$ and $g_s$ yielding a consistent DGP effective theory, where both the DGP crossover scale and the curvature radii are longer than the cutoff.
There is one dimensionless parameter, $N g_s$, controlling both how much longer than the cutoff $L_{\rm DGP}$ is, and how much {\em shorter} than the cutoff the radius of curvature is.
The results of this paper suggest that this obstruction may be generic and model-independent. In other words, it may be that the parameter space of DGP-like
models that can be obtained from UV-complete theories is smaller than the 
space determined by demanding self-consistency of the effective field
theory. Within that subspace, we expect to find no solution with pathologies
like superluminal signal propagation.

\subsection*{Acknowledgments}
\ \ \ \ \
We would like to thank Gregory Gabadadze and Matt Kleban
for useful comments. AN would also like to thank Sergei Dubovsky for earlier collaboration on this subject.  KH would like to thank Claire Zukowski for reading the manuscript.  KH is supported in part by by DOE grant DE-FG02-92ER40699.  MP is supported in part by NSF grants
PHY-0245068 and PHY-0758032, and by the 
ERC Advanced Investigator Grant n.226455 
{\em Supersymmetry, Quantum Gravity and Gauge Fields (Superfields)}.
 MP would also 
like to thank the Department of Physics 
at Columbia University for its support and hospitality during his sabbatical
visit in the Fall semester, 2008. 



\end{document}